\documentclass[twocolumn,apj]{emulateapj}

\usepackage{graphicx}
\usepackage{amsmath}
\usepackage{array}
\usepackage{tabularx}

\usepackage{natbib}

\def\be{\begin{equation}}
\def\ee{\end{equation}}
\def\ba{\begin{eqnarray}}
\def\ea{\end{eqnarry}}
\def\bal#1\eal{\begin{align}#1\end{align}}

\newcolumntype{W}{>{\raggedright\arraybackslash}X}
\newcolumntype{Y}{>{\raggedleft\arraybackslash}X}
\newcolumntype{Z}{>{\centering\arraybackslash}X}

\begin{document}
\title{Compositionally-driven convection in the oceans of accreting neutron stars}
\author{Zach Medin and Andrew Cumming}
\affil{Department of Physics, McGill University, 3600 rue University, Montreal, QC H3A 2T8, Canada; zmedin@physics.mcgill.ca, cumming@physics.mcgill.ca}

\begin{abstract}
We discuss the effect of chemical separation as matter freezes at the
base of the ocean of an accreting neutron star, and argue that the
retention of light elements in the liquid acts as a source of buoyancy
that drives a slow but continual mixing of the ocean, enriching it
substantially in light elements, and leading to a relatively uniform
composition with depth. We first consider the timescales associated
with different processes that can redistribute elements in the ocean,
including convection, sedimentation, crystallization, and
diffusion. We then calculate the steady state structure of the ocean
of a neutron star for an illustrative model in which the accreted
hydrogen and helium burns to produce a mixture of O and Se. Even
though the H/He burning produces only 2\% oxygen by mass, the steady
state ocean has an oxygen abundance more than ten times larger, almost
40\% by mass. Furthermore, we show that the convective motions
transport heat inwards, with a flux of $\approx 0.2$ MeV per nucleon
for an O-Se ocean, heating the ocean and steepening the outwards
temperature gradient. The enrichment of light elements and heating of
the ocean due to compositionally-driven convection likely have
important implications for carbon ignition models of superbursts.
\end{abstract}


\section{Introduction}
\label{sec:intro}

The ocean of an accreting neutron star is composed of a variety of
elements with atomic number Z = 6 and larger, formed by nuclear
burning of the accreted hydrogen and helium at low densities. The term
ocean refers to the fact that the Coulomb interaction energy between
ions is greater than the thermal energy, such that the ions behave
like a liquid. The ocean is of interest as the site of long duration
thermonuclear flashes such as superbursts (Cumming \& Bildsten 2001;
Strohmayer \& Brown 2002; Kuulkers 2004) and intermediate duration
bursts (in~'t Zand et al. 2005; Cumming et al. 2006), non-radial
oscillations (Bildsten \& Cutler 1995; Piro \& Bildsten 2005), and
because the matter in the ocean eventually solidifies as it is
compressed to higher densities by ongoing accretion, and so determines
the thermal, mechanical and compositional properties of the neutron
star crust (Haensel \& Zdunik 1990; Brown \& Bildsten 1998; Schatz et
al. 1999).

At the base of the ocean, matter freezes as it is compressed by
continuing accretion, becoming part of the solid crust. Horowitz et
al. (2007) carried out molecular dynamics simulations of the freezing
of a mixture of 17 species taken from a calculation of rp-process
hydrogen and helium burning and hence representative of the kind of
mixture expected to make up the ocean of an accreting neutron star
(Schatz et al. 2001; Gupta et al. 2007). They found that this mixture
underwent chemical separation during crystallization, such that light
elements (charge number $Z \alt 20$) were preferentially left behind
in the liquid, whereas heavier elements were preferentially
incorporated into the solid. In a previous paper (Medin \& Cumming
2010, hereafter Paper I) we showed that this result can be understood
by generalizing previous work using fits to the free energies of the
liquid and solid states of binary and tertiary plasmas.

In this paper, we address the implications of chemical separation for
the structure and composition of the ocean. Horowitz et al. (2007)
raised the question of what the steady-state ocean would look like,
since the matter entering the crust is enriched in certain elements
compared to others, and therefore different from the mean ocean
composition. We investigate this question here, and argue that the
retention of light elements in the liquid acts as a source of buoyancy
that drives a slow but continual mixing of the ocean, enriching it
substantially in light elements and leading to a relatively uniform
composition with depth. The steady state arises as the ocean enriches
in light elements to the point where the composition of the solid that
forms upon freezing matches the composition of matter entering the top
of the ocean.

One motivation for studying this problem comes from models for
superbursts which involve thermally-unstable carbon burning in the
deep ocean of the neutron star (Cumming \& Bildsten 2001; Strohmayer
\& Brown 2002). The energy release in these very long duration
thermonuclear flashes, inferred from fitting their lightcurves
(Cumming et al. 2006), corresponds to carbon fractions of $\approx
20\%$. This has been challenging to produce in models of the nuclear
burning of the accreted hydrogen and helium. If the hydrogen and
helium burn unstably, the amount of carbon produced is $\alt 1\%$
(Woosley et al. 2004), and whereas stable burning can produce large
carbon fractions (Schatz et al. 2003), time-dependent models do not
show stable burning at the $\approx 10\%$ Eddington accretion rates of
superburst sources (although observationally, superburst sources show
evidence that much of the accreted material may not burn in Type I
bursts; in~'t Zand et al. 2003).

Perhaps even more problematic than making enough carbon is that carbon
ignition models for superbursts require large ocean temperatures
$\approx 6\times10^8~{\rm K}$ at the ignition depth, which are
difficult to achieve in standard models of crust heating (e.g.,
Cumming et al. 2006; Keek et al. 2008). Similarly, Brown \& Cumming
(2009) inferred a large inwards heat flux in the outer crust of the
transiently-accreting neutron stars MXB 1659-29 and KS 1731-260 by
fitting their cooling curves in quiescence. Both of these observations
imply an additional heating source in the outer crust or ocean is
needed.  In this paper, we begin to address the question of to what
extent chemical separation could enrich the ocean in carbon and other
light elements, or provide a heat source that could alleviate some of
the difficulty of matching the observations of superbursts and
transient cooling.

We begin in \S\ref{sec:mixing} by reviewing the physics of chemical
separation, and discussing the timescales on which accretion,
crystallization, diffusion, sedimentation, and convection occur,
leading us to a picture of compositionally-driven convection. In
\S\ref{sec:convect} we calculate the structure of the steady-state
ocean for two simplified models: first, accretion of a two-component
mixture composed of Se and either O or Fe; and second, accretion of a
mixture of H and He which then burns to produce these heavier-element
mixtures. In \S\ref{sec:therm} we consider the effect of the mixing on
the thermal profile, and calculate the heating of the ocean due to the
convective transport of light elements outwards. Finally, in
\S\ref{sec:discuss} we discuss the implications of our results.

\begin{figure}
\begin{center}
\includegraphics[width=\columnwidth]{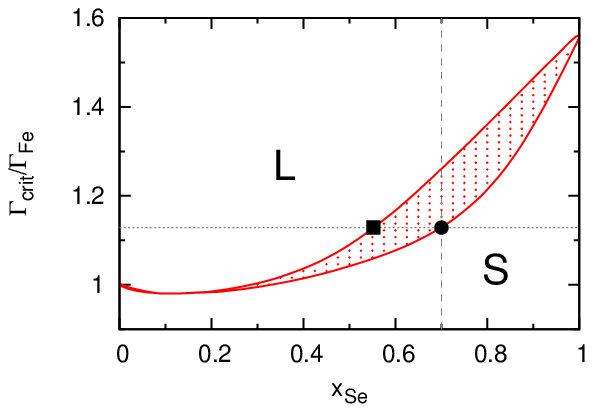}
\includegraphics[width=\columnwidth]{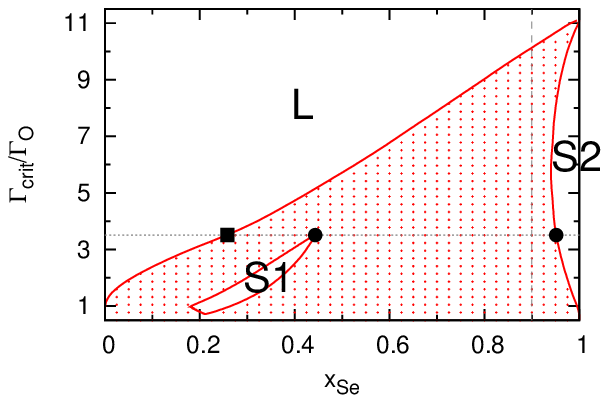}
\end{center}
\caption[The phase diagram]
{(Color online) The phase diagram for crystallization of an
${}^{56}$Fe-${}^{79}$Se mixture (top panel) and an
${}^{16}$O-${}^{79}$Se mixture (bottom panel) in a $T = 3\times10^8$~K
ocean. The Coulomb coupling constants $\Gamma_{\rm Fe,O}$ are given in
terms of $\Gamma_{\rm crit} \approx 175$, the value at which a
single-species plasma crystallizes. The stable liquid region of each
phase diagram is labeled `L', the stable solid region(s) are labeled
`S' or `S1' and `S2', and the unstable region is filled with plus
symbols. Additionally, in each panel the composition at the top of the
ocean is marked by a vertical dashed line, the ocean-crust boundary is
marked by a horizontal dotted line, the composition of the liquid at
the base of the ocean is marked by a filled square, and the
composition of the solid(s) in the outer crust are marked by filled
circles.}
\label{fig:phase}
\end{figure}

\section{Mixing processes in the ocean}
\label{sec:mixing}

\subsection{Phase diagrams and chemical separation}

The degree of chemical separation on freezing can be understood from
the phase diagram for the mixture. Figure~\ref{fig:phase} shows two
examples of phase diagrams, calculated as described in Paper I. The
upper panel is for a mixture of ${}^{56}$Fe ($Z=26$) and ${}^{79}$Se
($Z=34$), the lower panel for a mixture of ${}^{16}$O ($Z=8$) and
${}^{79}$Se ($Z=34$). In the first case the diagram is of azeotrope
type; in the second case, for which the ratio of atomic numbers $Z$ is
greater, the diagram is a more complicated eutectic type. In each
case, the x-axis shows the number fraction of Se and the y-axis shows
the inverse Coulomb coupling parameter for the light species (Fe or O
respectively) $\Gamma^{-1}$. The Coulomb coupling parameter for
species $i$ is
\bal
\Gamma_i = {}& \frac{Z_i^{5/3} e^2}{k_BT} \left(\frac{4\pi\rho Y_e}{3m_p}\right)^{1/3} \nonumber\\
 = {}& 204~\rho_9^{1/3}\left(\frac{T_8}{3}\right)^{-1}\left(\frac{Z_i}{34}\right)^{5/3}\left(\frac{Y_e}{0.43}\right)^{1/3} \,,
\label{eq:gamma}
\eal
where $Z_i$ is the charge of the ion, $Y_e = \langle Z \rangle/\langle
A \rangle$, $\langle Z \rangle$ and $\langle A \rangle$ are the
average charge and mass per ion of the mixture, $\rho_9 =
\rho/(10^9~{\rm g/cm^3})$ the density, and $T_8 = T/(10^8~{\rm K})$ the
temperature. For a single species of ion, solidification occurs when
$\Gamma_1 > \Gamma_m \approx 175$ (e.g., Potekhin \& Chabrier 2000). Note
that for a given $\rho$ and $T$ (or depth in the star), $\Gamma_i$ is
nearly constant with composition. As a fluid element is compressed by
accretion to higher density, $\Gamma_1$ increases, moving down in the
phase diagram. The shaded regions represent unstable regions of the
phase diagram. A fluid element with composition and $\Gamma_1$ that
lies inside the unstable region will undergo phase separation,
separating into two phases with compositions on each side of the
unstable region. In this way, chemical separation occurs. Note that
the curves that bound the unstable region are commonly referred to as
the liquidus and solidus curves, respectively. A liquid with
composition of the liquidus curve at a given $\Gamma_1$ is in
equilibrium with the solid which has the composition of the solidus
curve at that $\Gamma_1$.

Now consider a particular mixture of Fe and Se entering the top of the
ocean with $x_{\rm Se} = 0.7$ (this corresponds to 77\% Se and 23\% Fe
by mass), indicated by the vertical dashed line in the upper panel of
Fig.~\ref{fig:phase}. In steady state, the solid forming at the top of
the crust must have this same composition, so that the freezing point
must lie at $\Gamma_m/\Gamma_{\rm Fe} \approx 1.1$ as marked by the
filled circle in Fig.~\ref{fig:phase}. The corresponding liquid
composition in equilibrium with the solid at this $\Gamma_{\rm Fe}$ is
indicated by the solid square. The phase diagram shows that the liquid
at the base of the ocean must have a composition $x_{\rm Se} \approx
0.56$ (63\% Se by mass) in order to make the solid composition
demanded by steady state.

The lower panel of Fig.~\ref{fig:phase} shows a more complicated
example. The vertical dotted line marks an incoming composition with
$x_{\rm Se} = 0.9$ (corresponding to 98\% Se and 2\% O by mass). In
this case, there is no single solid phase with this
composition. Instead, at $\Gamma_m/\Gamma_{\rm O} \approx 3.5$, a
mixture of two solid phases indicated by the filled circles forms, and
the liquid at the base of the ocean has $x_{\rm Se} \approx 0.25$
(63\% Se by mass). Again, in order to reach steady state, the base of
the ocean must adjust its composition until it is significantly
enriched in light elements compared to the composition at the top of
the ocean.

\subsection{Crystallization of solid particles}

To see how the ocean is able to adjust its composition profile to
achieve steady state, we first note that solid particles crystallize
and sediment out rapidly compared to the accretion timescale on which
matter is compressed. The accretion time is $t_{\rm accr} = H_P/v_{\rm
accr} = y/\dot{m}$ or
\be
t_{\rm accr} = 3.2~\frac{y_{12}}{\dot{m}_4}~{\rm yrs} = 7.2~\frac{\rho_9^{4/3}}{\dot{m}_4}\left(\frac{Y_e}{0.43}\right)^{4/3}\left(\frac{g_{14}}{2.45}\right)^{-1}~{\rm yrs,}
\label{eq:taccr}
\ee
where $H_P = P/\rho g = y/\rho$ is the pressure scale height, $y$ is
the column depth, $\dot{m}$ is the local accretion rate per unit area
and in the last step we use the fact that the pressure in the ocean is
dominated by relativistic degenerate electrons. We scale to a typical
rate $\dot{m}_4 = \dot{m}/(10^4~{\rm g~cm^{-2}~s^{-1}})$ (to set the
scale note that the Eddington rate is $\dot{m}_{\rm Edd} \approx
10^5~{\rm g~cm^{-2}~s^{-1}}$). We choose the gravity $g_{14} =
g/(10^{14}~{\rm cm/s^2}) = 2.45$ corresponding to a $1.4~M_\odot$, $R
= 10~{\rm km}$ neutron star.

First, consider crystallization. The nucleation rate, the rate at
which solid clusters of a size large enough to be stable are formed,
is not currently well understood. Theory typically predicts nucleation
rates orders of magnitude smaller than experiment or simulations
indicate (see Vehkam\"aki 2006). For an OCP for example, the
nucleation rate at $\Gamma_1 = 300$ from the theoretical model of
Ichimaru et al. (1983) is $\sim 10^{-6}$ of its value from the
simulation of Daligault (2006b). Cooper \& Bildsten (2008) recently
derived a value within a factor of 10 of the simulations, but their
formulation predicts the formation of stable solid clusters even at
$\Gamma_1 < 175$ when the liquid phase should be absolutely stable
with respect to the solid. Taking the Ichimaru et al. (1983) results,
we find that the amount of undercooling necessary for the nucleation
rate to become comparable to the accretion rate is $\Gamma_1 -
\Gamma_m \sim 0.1\Gamma_m$, but given that the theory underpredicts
the simulation results, the amount of undercooling required is
probably much less than this. Additionally, at the base of the ocean
solid clusters do not need to wait for nucleation sites to form but
can crystallize on the existing crust.

Once a cluster forms it increases in size at the crystallization
velocity which is of order $v_{\rm crys} \sim a\omega_p$ (e.g.,
Kelton, Greer, \& Thompson 1983), where $a$ is the mean ion spacing
and $\omega_p$ is the ion plasma frequency given by $\omega_p =
\sqrt{4\pi \rho} (Y_ee/m_p) = 1.4\times10^{19}
\rho_{9}^{1/2}(Y_e/0.43)~{\rm rad/s}$. In a multicomponent plasma
crystal growth is slower, since as a solid cluster grows chemical
separation means that the liquid surrounding the cluster is depleted
more and more of the particles necessary to form the solid. The
crystallization rate therefore depends on the rate at which diffusion
can replenish the depleted particles. In a liquid with $100 < \Gamma <
300$, where $\Gamma = \langle Z^{5/3} \rangle\Gamma_i/Z_i^{5/3}$ is
the {\it average} Coulomb coupling constant, the diffusion coefficient
for species $i$ is
\be
D \approx 8 \left(\frac{\langle Z \rangle}{Z_i}\right)^{2/3} \omega_pa^2\Gamma^{-2.5}~{\rm cm^2/s}
\label{eq:Di}
\ee
(Horowitz et al. 2010; see also Daligault \& Murillo 2005). We assume
that to bind to a cluster, new material must travel a distance $l$
equal to the current size of the cluster; in this case, the cluster
growth time is $l^2/D \approx 10^{-4}
\omega_p^{-1}(l/a)^2(\Gamma/100)^{2.5}$. For a solid cluster of $N_s$
particles, $(l/a)^2 \approx N_s^{2/3}$.

Therefore as soon as the mixture encounters the liquidus line in the
phase diagram, we expect solid particles to rapidly form and
grow. Once formed, the solid particles will quickly sediment out. We
estimate the sedimentation velocity for solid clusters following
Bildsten \& Hall (2001) and Brown et al. (2002). The sedimentation
velocity is given by the Einstein relation
\be
v_{\rm sed} \simeq \frac{D}{k_BT} A_sN_s m_pg\frac{\Delta Y_e}{Y_e} \,,
\label{eq:vsed}
\ee
where $\Delta Y_e/Y_e = 1 - Y_{e,s}/Y_{e,l}$ is the contrast between
$Y_e$ for the solid particles and the background fluid.  Note that in
Eq.~(\ref{eq:vsed}) we have neglected the contribution of the ions to
the buoyancy force. Usually this amounts to a $\sim 10\%$ correction;
but when $\Delta Y_e = 0$ this is the dominant term (Mochkovitch
1983). We use the Stokes-Einstein relation to estimate the mobility
$D/(k_BT) = \left(4\pi\eta aN_s^{1/3}\right)^{-1}$, where the OCP shear
viscosity is $\eta \approx 0.2\rho\omega_pa^2(\Gamma/100)^{1.5}~{\rm
g~cm^{-1}~s^{-1}}$ (Donk\'o \& Ny\'iri 2000; Daligault 2006a). We
find
\be
v_{\rm sed} \agt 3\times10^{-7} \rho_9^{-0.6} \left(\frac{T_8}{3}\right)^{0.3} \left(\frac{\Delta Y_e/Y_e}{0.01}\right) N_s^{2/3}~{\rm cm/s} \,.
\ee
Comparing with the accretion velocity $v_{\rm accr} = \dot{m}/\rho =
10^{-5}\dot{m}_4\rho_9^{-1}~{\rm cm/s}$, we find that the critical
cluster size above which the particles sediment out is
\be
N_{s,\rm crit} = 200~\rho_9^{-0.6} \left(\frac{T_8}{3}\right)^{-0.45} \dot{m}_4^{3/2} \left(\frac{\Delta Y_e/Y_e}{0.01}\right)^{-3/2} \,.
\label{eq:Nscrit}
\ee
Note that the simple nature of our estimates means that there are
considerable uncertainties in the crystallization and sedimentation
velocities we find here. Despite this, however, the timescale for a
cluster to grow to a size $N_{s,\rm crit}$ is so short that the
conclusion that the solid particles rapidly fall out of the ocean
seems inescapable.

Setting $N_{s,\rm crit} = 1$ in Eq.~(\ref{eq:Nscrit}) provides an
estimate of when relative separation of light and heavy elements
(without forming solid clusters) is expected to occur. This gives
$\dot{m} = 300~{\rm g~cm^{-2}~s^{-1}} \approx 0.003 \dot{m}_{\rm
Edd}$, below the accretion rates of persistent LMXBs or most
transient LMXBs in outburst. However, separation of light and heavy
elements is something that should be considered during quiescent
periods in transient accretors (Brown et al. 2002) or at low accretion
rates (Peng, Brown, \& Truran 2007).

\subsection{Compositional buoyancy and convection}
\label{sub:buoy}

After the solid particles form and sediment out, the fluid left behind
is lighter than the fluid immediately above it and so will have a
tendency to buoyantly rise. This is counteracted by the thermal
profile, which is stably stratified in the absence of a composition
gradient such that a rising fluid element will be colder than its
surroundings and will tend to sink back down. A measure of the
buoyancy is the convective discriminant ${\mathcal A}$, which is
related to the Brunt-V\"ais\"al\"a frequency $N^2=-g{\mathcal A}$ (Cox
1980). For a two-component mixture in the ocean, we can write
(Bildsten \& Cumming 1998)
\be
{\mathcal A}H_P = \frac{\chi_T}{\chi_\rho}\left(\nabla -\nabla_{\rm ad}\right)+\frac{\chi_{X}}{\chi_\rho}\nabla_{X} \,;
\label{eq:A}
\ee
here, $X$ is the mass fraction of the lighter element (Fe or O in
the examples above), $\chi_X = \partial \ln P/\partial \ln
X|_{T,\rho}$, $\chi_\rho = \partial \ln P/\partial \ln \rho|_{X,T}$,
$\chi_T = \partial \ln P/\partial \ln T|_{X,\rho}$, and the
temperature and composition gradients are $\nabla = -H_P(d\ln T/dr)$
and $\nabla_X = -H_P(d\ln X/dr)$. The adiabatic gradient is taken at
constant entropy $S$ and composition: $\nabla_{\rm ad} = -H_P(d\ln T
/dr|_{S,X})$. Note that $\chi_X$, $\chi_T$, and $\chi_\rho$ are all
positive quantities. If ${\mathcal A} < 0$ or $N^2 > 0$ the ocean is
stable to convection. For example, if the composition is uniform so
that $\nabla_X = 0$, stability to convection requires the familiar
condition $\nabla < \nabla_{\rm ad}$. Only one term describing the
variation of the composition is needed in Eq.~(\ref{eq:A}) since we
consider a two-component mixture. We generalize to more than two
species in the next section.

In steady state the composition profile is lighter with increasing
depth: $\nabla_X > 0$. Such a profile will not lead to convection as
long as the gradient is small enough that the destabilizing effect of
the composition profile is compensated by the thermal buoyancy
represented by the first term in Eq.~(\ref{eq:A}). The maximum stable
composition gradient is
\be
\nabla_{X,\rm max} = \frac{\chi_T}{\chi_X}\left(\nabla_{\rm ad}-\nabla\right) \approx \frac{\chi_T\nabla_{\rm ad}}{\chi_X} \,,
\label{eq:nablaX1}
\ee
where we assume that the large thermal conductivity in the ocean due
to the degenerate electrons results in an almost isothermal profile
$\nabla \ll \nabla_{\rm ad}$ (see \S\ref{sec:therm}).

As accretion continues, light elements are continually deposited at
the base of the ocean and must be transported upwards by
convection. We expect therefore that the composition gradient will
adjust to be close to but slight greater than $\nabla_{X,\rm max}$ so
as to result in the required convective flux of composition $F_X =
\dot{m} (X_0-X_b)$, where $X_0$ is the incoming composition and
$X_b$ is the composition at the base of the ocean (that results in
freezing of solid with mass fraction $X_0$).

We can estimate $\nabla_X - \nabla_{X,\rm max}$ in steady state and
the corresponding convective velocity using mixing length theory. The
acceleration of a fluid element is $g{\mathcal A}$, giving a
convective velocity
\be
v_{\rm conv}^2 \sim g l_m^2 \frac{\chi_X\nabla_X-\chi_T\nabla_{\rm ad}}{H_P\chi_\rho} \,,
\label{eq:vconvanal}
\ee
where $l_m$ is the mixing length and we again assume $\nabla = 0$ for
simplicity. After moving a distance $l_m$, the mass fraction differs
from its surroundings by an amount $(l_m/H_P)X\nabla_X$, implying that
there is a flux of composition
\be
F_X \sim \rho v_{\rm conv} \frac{l_m X\nabla_X}{H_P} \,.
\ee
Setting this equal to the steady state flux at the base of the ocean
$F_X = \dot{m} (X_0-X_b) = \rho v_{\rm accr} (X_0-X_b)$ gives the
maximum convective velocity
\bal
v_{\rm conv} \sim {}& v_{\rm accr} \left(\frac{H_P}{l_m}\right) \frac{X_0-X_b}{X_b\nabla_{X,\rm max}} \nonumber\\
 \sim {}& v_{\rm accr} \left(\frac{H_P}{l_m}\right) \left(\frac{X_0-X_b}{X_b}\right) \nabla_{\rm ad}^{-1} \frac{\chi_X}{\chi_T} \,.
\label{eq:vconvmax}
\eal
In the deep ocean where pressure is dominated by degenerate electrons,
all factors on the right hand side of Eq.~(\ref{eq:vconvmax}) are of
order unity except for $\chi_X \approx \chi_\rho (1-Y_{\rm Se}/Y_e)
\sim 0.1$, where $Y_{\rm Se} = 34/79$; and $\chi_T \approx
(T/P_e)(\partial P_i/\partial T) \sim 10k_BT/[\langle Z \rangle E_F]$
(e.g., Hansen \& Kawaler 1994), where $P_e$ is the electron pressure,
$P_i$ the ion pressure, and $E_F$ is the Fermi energy excluding rest
mass. For an ocean temperature of $3\times10^8~{\rm K}$ and $E_F = 5.1
\rho_9^{1/3}Y_e^{1/3}~{\rm MeV}$ (the electrons are relativistic),
$\chi_T \sim 0.001$. Therefore, $v_{\rm conv}$ is about two orders of
magnitude larger than the accretion velocity. Given that chemical
separation is being driven by accretion, it may be surprising that
$v_{\rm conv}$ is much larger than $v_{\rm accr}$. The reason is that
only a very shallow composition gradient can be tolerated in the deep
ocean where $k_BT \ll E_F$, requiring $v_{\rm conv} \gg v_{\rm accr}$
to transport the required flux of composition.

Comparing Eqs.~(\ref{eq:vconvanal}) and (\ref{eq:vconvmax}), we see
that $\nabla_X-\nabla_{X,\rm max} \alt (v_{\rm accr}/c_s)^2
\left[\chi_\rho\chi_X (X_0-X_b)^2/\nabla_{\rm
ad}^2\chi_T^2X_b^2)\right] \ll 1$, where $c_s = (gH_P)^{1/2}$ is
the sound speed. In other words the convective velocities needed to
transport the flux of light elements through the ocean are very
subsonic, implying the composition gradient is extremely close to the
marginally stable gradient $\nabla_X \approx \nabla_{X,\rm max}$. This
is analogous to efficient convective heat transport for which $\nabla
\approx \nabla_{\rm ad}$.

Another piece of physics that could potentially play a role is
diffusion. Inwards diffusion of composition will occur down the
composition gradient $\nabla_X$, and in principle if efficient enough
could mediate the need for convection. The diffusive flux is much
smaller than the convective flux, however. To see this, we use
Eq.~(\ref{eq:Di}) for a selenium OCP:
\be
D = 2\times10^{-6} \rho_9^{-1} \left(\frac{T_8}{3}\right)^{2.5}~{\rm cm^2/s} \,.
\ee
The diffusion time across a composition scale height $l_X =
H_P/\nabla_X$, where the pressure scale height $H_P = 1900
\rho_9^{1/3}(Ye/0.43)^{4/3}(g_{14}/2.45)^{-1}~{\rm cm}$ is
\be
t_D = \frac{l_X^2}{D} = 5\times10^4 \nabla_X^{-2}\rho_9^{1.3} \left(\frac{T_8}{3}\right)^{-2.5}~{\rm yrs.}
\ee
This is much longer than the accretion timescale given by
Eq.~(\ref{eq:taccr}), or the convective turnover time given by
\be
t_{\rm conv} = l_m/v_{\rm conv} \sim 100~l_m/v_{\rm accr} \,,
\label{eq:tconv}
\ee
implying that microscopic diffusion does not play a significant role
in transporting light elements across the ocean.

We also note that our conclusion that the ocean is convectively
unstable depends on the slope of the liquidus curve in the phase
diagram. In all of the cases we have considered, the composition
profile that would exist if the ocean followed the liquidus curve as
fluid elements move to higher pressure is too steep to be maintained
by thermal buoyancy. If this were not the case, a different picture
would result: the composition in the ocean at a given pressure would
correspond to the composition of the liquidus curve at each depth, and
a steady hail of solid particles would fall through the ocean to the
solid crust at the base. Given the phase diagrams for two- and
three-component plasmas we have calculated, however, this situation
appears not to arise.

We therefore expect that the ocean adjusts its composition from $X_0$
at the top to $X_b$ at the base with a mixing zone in which the
gradient is $\nabla_X \approx \nabla_{X,\rm max}$. Because this
gradient is very shallow in the deep degenerate part of the ocean
($\nabla_{X,\rm max} \sim 10k_BT/[\langle Z \rangle E_F]$), we expect
the mixing zone to have substantial thickness. We investigate this in
the next section with detailed models of the steady-state ocean, and
find that the entire ocean is expected to undergo mixing all the way
up to the layer of light elements that supplies the ocean with new
material through nuclear burning. The shallow gradient also means that
there is a nearly uniform composition throughout the bulk of the
ocean, which is set by the phase diagram at the freezing point.

\section{Illustrative models of a convective ocean}
\label{sec:convect}

In \S\ref{sec:mixing} we argued that crystallization and sedimentation
would drive a convective instability at the base of the ocean. We now
calculate detailed but illustrative models of the steady-state ocean
based on this picture. We use mixing length theory to calculate the
convective velocity, keeping the mixing length as a free parameter;
whether mixing length theory is appropriate for compositionally-driven
convection is an open question. As we discuss further in
\S\ref{sec:discuss}, we also neglect other hydrodynamical circulations
(e.g. Eddington-Sweet circulation driven by rotation), the effects of
rotation or magnetic fields on the convection itself, or other
transport mechanisms such as turbulent mixing that could lead to
transport of composition through the ocean.

We first allow the ocean to extend to arbitrarily low density; i.e.,
we first neglect the light element layer which overlies the ocean
(\S\ref{sub:2cp}) and then show how the light element layer can be
incorporated self-consistently into a steady-state model of the ocean
(\S\ref{sub:3cp}). The models in \S\ref{sub:2cp} and \S\ref{sub:3cp}
are isothermal; in \S\ref{sec:therm} we consider the effect of the
mixing on the thermal profile.

\subsection{A first model of the ocean}
\label{sub:2cp}

We first calculate the steady state in the ocean for the case where
nuclear reactions are unimportant in the convection zone. We show
results for the specific two-component mixtures discussed in
\S\ref{sec:mixing}, but for generality we keep the number of species
arbitrary in the derivations that follow. The continuity equation for
the flow of species $i$ is
\be
\frac{dX_i}{dt} + \mathbf{v}_{\rm accr} \cdot \nabla X_i = -\frac{1}{\rho} \nabla \cdot \left(\rho \mathbf{v}_{\rm conv} DX_i \right) \,,
\ee
where
\be
DX_i = \frac{\xi}{2} X_i\nabla_{X_i}
\ee
for an inward-convecting blob, and the convective velocity from mixing
length theory is given by (e.g., Kippenhahn \& Weigert 1994)
\be
v_{\rm conv}^2 = \xi^2\frac{gH_P}{8\chi_\rho}\left(\chi_T(\nabla - \nabla_{\rm ad}) + \sum_{i=1}^{n-1} \chi_{X_i}\nabla_{X_i}\right) \,,
\label{eq:vconv}
\ee
where $n$ is the total number of chemical species in the ocean, $X_i$
is the mass fraction of species $i$, and the parameter $\xi = l_m/H_P$
is the ratio between the convection mixing length $l_m$ and the scale
height. In mixing length theory the value of $\xi$ is highly
uncertain; we assume that $\xi = 1$ in this paper, but $\xi$ could be
an order of magnitude or two smaller than this.

In steady state and for vertical flow in plane-parallel geometry (a
good assumption since the ocean is thin compared to the stellar
radius), we have
\be
v_{\rm accr} \frac{dX_i}{dr} = \frac{\xi}{2\rho} \frac{d}{dr}\left(\rho v_{\rm conv} X_i\nabla_{X_i}\right) \,.
\ee
Integrating from the top of the ocean to some depth within the ocean,
using the fact that $\rho v_{\rm accr} = \dot{m}$ is a constant, we
find
\be
v_{\rm accr} (X_i-X_{i,0}) = v_{\rm conv} \frac{\xi}{2} X_i\nabla_{X_i} \,.
\label{eq:vconv2}
\ee
Note that at the top of the convection zone, when $v_{\rm conv} = 0$,
we have $X_i = X_{i,0}$ for each $X_i$. Rewriting
Eq.~(\ref{eq:vconv2}) gives
\be
\frac{dX_i}{d\ln P} = \frac{v_{\rm accr}}{v_{\rm conv}} \frac{2}{\xi} (X_i-X_{i,0})
\label{eq:dXdP0}
\ee
which can be integrated for each species $i=1,2,\ldots,n-1$ (the
composition of species $n$ follows from the constraint $\sum_{i=1}^n
X_i=1$).

The $n-1$ differential equations (\ref{eq:dXdP0}) are coupled because
the convective velocity depends on a sum over all species. To obtain
an expression for the convective velocity, we multiply
Eq.~(\ref{eq:vconv2}) by $\chi_{X_i}$ and sum over species to obtain
\be
\sum_{i=1}^{n-1} \chi_{X_i} \nabla_{X_i} = \frac{v_{\rm accr}}{v_{\rm conv}} \frac{2}{\xi} \sum_{i=1}^{n-1} \chi_{X_i}\frac{X_i-X_{i,0}}{X_i}.
\label{eq:step1}
\ee
We argued in \S\ref{sec:mixing} that small convective velocities
$v_{\rm conv} \ll c_s$ are required to transport the composition
flux. Therefore,
\be
\sum_{i=1}^{n-1} \chi_{X_i} \nabla_{X_i} \approx \chi_T (\nabla_{\rm ad}-\nabla)
\label{eq:nablaX2}
\ee
[cf. Eq.~(\ref{eq:nablaX1})], and we can replace the left-hand side of
Eq.~(\ref{eq:step1}) with $\chi_T (\nabla_{\rm ad} -
\nabla)$. Rearranging, we find an expression for $v_{\rm conv}$:
\be
v_{\rm conv} = v_{\rm accr} \frac{2}{\xi} \frac{1}{\chi_T(\nabla_{\rm ad}-\nabla)} \sum_{i=1}^{n-1} \chi_{X_i}\frac{X_i-X_{i,0}}{X_i} \,.
\label{eq:vconv3}
\ee
As a check we see that for two species ($n = 1$),
Eqs.~(\ref{eq:dXdP0}) and (\ref{eq:vconv3}) give $d\ln X/d\ln P =
(\chi_T/\chi_X)(\nabla_{\rm ad}-\nabla)$ which is the marginally
stable composition gradient (\S\ref{sub:buoy}).

\begin{figure}
\begin{center}
\includegraphics[width=\columnwidth]{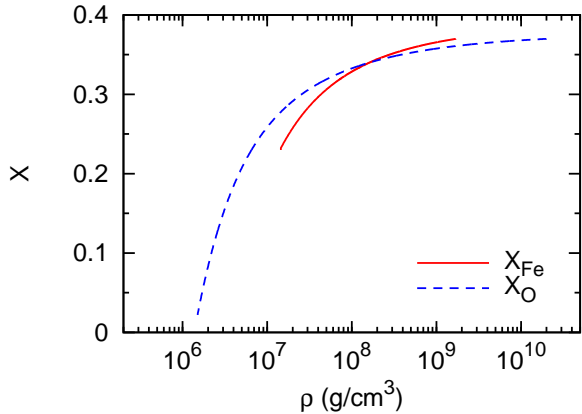}
\end{center}
\caption[The composition profile, no H/He burning layer]
{(Color online) The composition profile in the convection zone of a $T
= 3\times10^8$~K ocean composed of ${}^{56}$Fe, ${}^{79}$Se (solid
line) and ${}^{16}$O, ${}^{79}$Se (dashed line). The hydrogen/helium
burning layer is neglected.}
\label{fig:tcp}
\end{figure}

To obtain a solution, the location and composition of the base of the
ocean are specified, and then Eq.~(\ref{eq:dXdP0}) is integrated
outward for each $X_i$. The top of the convection zone is located at
the point where the composition matches the incoming composition, $X_i
= X_{i,0}$. In Fig.~\ref{fig:tcp} we present our results for the
${}^{56}$Fe-${}^{79}$Se and ${}^{16}$O-${}^{79}$Se systems described
in \S\ref{sec:mixing}. The ocean temperature is taken to be $T =
3\times10^8~{\rm K}$. To calculate the various $\nabla$'s and
$\chi$'s, we assume that the electron pressure is given by the fitting
formula of Paczy\'nski (1983), and following Paper I we include
Coulomb corrections for the ions using the free energy from DeWitt \&
Slattery (2003). Note that the equations used in this section to
calculate the composition profile are all independent of $\xi$; only
the convective velocity depends on this value. Therefore, the
potentially large error introduced by our choice of $\xi$ does not
affect our results.

\begin{figure}
\begin{center}
\includegraphics[width=\columnwidth]{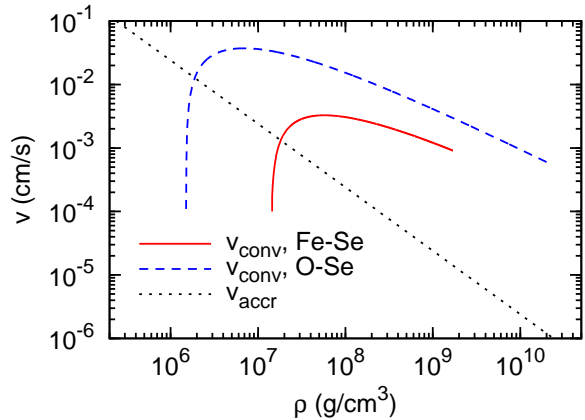}
\end{center}
\caption[Accretion and convective velocities]
{(Color online) The accretion velocity for $\dot{m}=3\times10^4~{\rm
g~cm^{-2}~s^{-1}}$, and the convective velocity within the convection
zone of a $T = 3\times10^8$~K ocean composed of ${}^{56}$Fe,
${}^{79}$Se (solid line) and ${}^{16}$O, ${}^{79}$Se (dashed line).}
\label{fig:vconv}
\end{figure}

The solutions show the expected behavior, that at large depths where
$k_BT \ll E_F$, the composition gradient is very shallow, but as the
integration continues further upwards, the gradient steepens as $E_F$
drops. In each case, the ocean is substantially enriched in light
elements throughout most of its mass. Figure~\ref{fig:vconv} compares
the convective velocity with the accretion velocity for these two
cases, assuming $\xi=1$ and $\dot{m}_4 = 3$ (i.e., an accretion rate
of $0.3\dot{m}_{\rm Edd}$). Towards the top of the ocean, where the
composition gradient becomes significant ($E_F$ becomes comparable to
$k_BT$), the convective velocity drops towards the accretion velocity.

\subsection{Including nuclear burning of light elements}
\label{sub:3cp}

Figure~\ref{fig:tcp} shows that in the oxygen-selenium two-species
model at $T = 3\times10^8~{\rm K}$, the convection zone extends into
the hydrogen/helium burning layer ($\rho \sim 10^5$--$10^7~{\rm
g/cm^3}$). This effect becomes more pronounced at lower
temperatures, as might be expected when $\dot{m} < 0.3\dot{m}_{\rm
Edd}$: empirically the density at the top of the convection zone
goes as $T^{2.3}$ for typical ocean temperatures, in both the O-Se and
the Fe-Se models. Therefore, an accurate model of the entire
convection zone must include the effects of nuclear reactions. As a
first approximation, we use the following simplified model for the
burning layer: At $\rho = 10^4~{\rm g/cm^3}$ the neutron star is
composed only of hydrogen and helium, with $X_{\rm H} = 0.7$ and
$X_{\rm He} = 0.3$. The relative abundance of hydrogen and helium is
maintained as these elements undergo nuclear reactions and convective
mixing, such that $X_{\rm H}/X_{\rm He} = 0.7/0.3$ is a constant
throughout the ocean. This mixture burns at the triple-alpha rate
(Hansen \& Kawaler 1994)
\be
R_{3\alpha} \approx 10^{-6} \left(\frac{\rho}{\rm g/cm^3}\right)^2 (0.3X_1)^2 \frac{e^{-44/T_8}}{T_8^3}~{\rm s^{-1}} \,,
\label{eq:R3a}
\ee
where $X_1$ is the mass fraction of the H/He composite with $X_{\rm H}
= 0.7X_1$ and $X_{\rm He} = 0.3X_1$. Each gram of the mixture burns to
$X_i^{\rm burn}$ grams of each chemical species $i>1$ (so that
$\sum_{i=2}^n X_i^{\rm burn}=1$).

We use this model for the burning layer to calculate the steady state
in the ocean for the case where nuclear reactions are important in the
convection zone. The continuity equation for each species becomes
\be
\frac{dX_1}{dt} + \mathbf{v}_{\rm accr} \cdot \nabla X_1 = -\frac{1}{\rho} \nabla \cdot \left(\rho \mathbf{v}_{\rm conv} DX_1 \right) - X_1 R_{3\alpha}
\ee
for $i=1$ and
\be
\frac{dX_i}{dt} + \mathbf{v}_{\rm accr} \cdot \nabla X_i = -\frac{1}{\rho} \nabla \cdot \left(\rho \mathbf{v}_{\rm conv} DX_i \right) + X_i^{\rm burn} X_1 R_{3\alpha}
\ee
for $i=2,3,\ldots,n$. In steady state we have
\be
v_{\rm accr} \frac{dX_1}{dr} = \frac{\xi}{2\rho} \frac{d}{dr}\left(\rho v_{\rm conv} X_1\nabla_{X_1}\right) + X_1 R_{3\alpha}
\ee
and
\be
v_{\rm accr} \frac{dX_i}{dr} = \frac{\xi}{2\rho} \frac{d}{dr}\left(\rho v_{\rm conv} X_i\nabla_{X_i}\right) - X_i^{\rm burn} X_1 R_{3\alpha} \,.
\ee
Above the convection zone (where $v_{\rm conv} = 0$) we use
\be
v_{\rm accr} \frac{dX_1}{d\ln P} = -X_1 R_{3\alpha} H_P
\label{eq:dXdPa1}
\ee
to solve for $X_1$ as a function of $P$, and then
\be
v_{\rm accr} \frac{dX_i}{d\ln P} = X_i^{\rm burn} X_1 R_{3\alpha} H_P
\ee
or
\be
X_i = (1-X_1)X_i^{\rm burn}
\label{eq:dXdPa2}
\ee
to find the $X_i$, $i > 1$ values. These equations are solved inward
from $\rho = 10^4~{\rm g/cm^3}$, where we set $X_1 = 1$ and $X_i = 0$
for $i > 1$. Within the convection zone we use the following method to
find the $X_i$'s: We first specify the location of the top of the
convection zone, $P_0$. The $X_{i,0}$ values used in the equations
below are then found by solving Eqs.~(\ref{eq:dXdPa1}) and
(\ref{eq:dXdPa2}) down to this point. The value of $P_0$ is not
known a priori, so the $X_i$'s are found for a given $P_0$ and then
the value is varied until the composition changes smoothly across the
convection zone boundary. We define a new variable, $W$, such that
\be
\frac{dW}{dr} = \frac{X_1 R_{3\alpha}}{v_{\rm accr}}
\label{eq:dwdr}
\ee
and $W = 0$ at the top of the convection zone (when $X_i =
X_{i,0}$). Then the equation for the convective velocity is
\bal
\sum_{i=1}^{n-1} \chi_{X_i}\frac{X_i-X_{i,0}}{X_i} - {}& W\left(\frac{\chi_{X_1}}{X_1} - \sum_{i=2}^{n-1} \frac{\chi_{X_i}}{X_i}X_i^{\rm burn}\right) \nonumber\\
 = {}& \frac{v_{\rm conv}}{v_{\rm accr}} \frac{\xi}{2} \sum_{i=1}^{n-1} \chi_{X_i} \nabla_{X_i} \,,
\eal
or
\bal
v_{\rm conv} = v_{\rm accr} \frac{2}{\xi} {}& \frac{1}{\chi_T(\nabla_{\rm ad} - \nabla)} \left[\sum_{i=1}^{n-1} \chi_{X_i}\frac{X_i-X_{i,0}}{X_i} \right. \nonumber\\
 {}& \left. - W\left(\frac{\chi_{X_1}}{X_1} - \sum_{i=2}^{n-1} \frac{\chi_{X_i}}{X_i}X_i^{\rm burn}\right)\right] \,.
\eal
The $n$ coupled differential equations to solve are
\be
\frac{dX_1}{d\ln P} = \frac{v_{\rm accr}}{v_{\rm conv}} \frac{2}{\xi} (X_1-X_{1,0}-W) \,;
\ee
\be
\frac{dX_i}{d\ln P} = \frac{v_{\rm accr}}{v_{\rm conv}} \frac{2}{\xi} (X_i-X_{i,0}-WX_i^{\rm burn}) \,,
\ee
for $i=2,3,\ldots,n-1$; and Eq.~(\ref{eq:dwdr}). We solve these
equations through iteration, first guessing the $X_i$ values at every
point in the convection zone, then using these values to find $W$, and
so on, until convergence is reached.

\begin{figure}
\begin{center}
\includegraphics[width=\columnwidth]{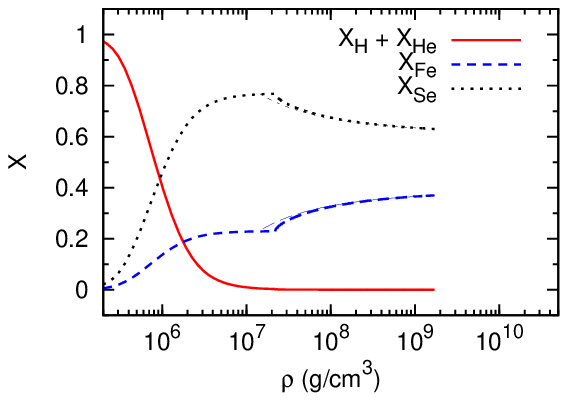}
\includegraphics[width=\columnwidth]{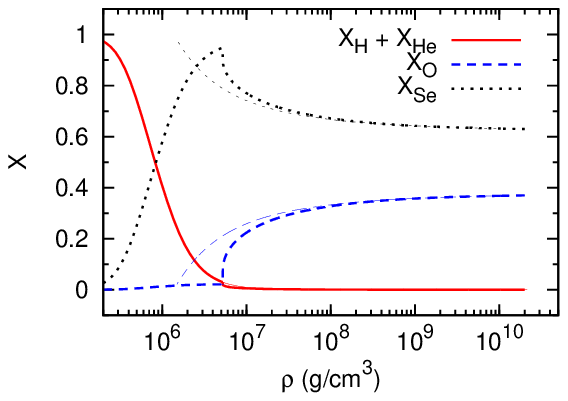}
\end{center}
\caption[The composition profile, with a H/He burning layer]
{(Color online) The composition profile within and directly above the
convection zone of a $T = 3\times10^8$~K ocean composed of
${}^{56}$Fe, ${}^{79}$Se (top panel) and ${}^{16}$O, ${}^{79}$Se
(bottom panel). A layer of hydrogen and helium is placed on top of the
ocean and nuclear reactions are crudely modeled. For comparison the
$X_{\rm Fe}$, $X_{\rm O}$, and $X_{\rm Se}$ profiles when the burning
layer is neglected (Fig.~\ref{fig:tcp}), and the $X_{\rm H} + X_{\rm
He}$ profiles when convection is turned off [Eqs.~(\ref{eq:dXdPa1})
and (\ref{eq:dXdPa2})] are also plotted, as thin lines with the same
patterns as the profiles from the full calculation. The convection
zone begins at the point where the two $X_{\rm H} + X_{\rm He}$ curves
diverge, at a density of $7.2\times10^6~{\rm g/cm^3}$ for the F-Se
system and $2.6\times10^6~{\rm g/cm^3}$ for the O-Se system.}
\label{fig:mcp}
\end{figure}

In Fig.~\ref{fig:mcp} we present our results for the
${}^{56}$Fe-${}^{79}$Se and ${}^{16}$O-${}^{79}$Se systems discussed
in \S\ref{sec:mixing}; again we choose $\dot{m}_4 = 3$, $\xi = 1$, and
an ocean temperature $T = 3\times10^8~{\rm K}$. For comparison we also
present the $X_{\rm Fe}$, $X_{\rm O}$, and $X_{\rm Se}$ profiles when
the hydrogen/helium burning layer is neglected
(\S\ref{sub:2cp}). These profiles are nearly the same as the profiles
when the burning layer is included, except at the top of the
convection zone where the latter profiles are very steep. The change
in the slope of the profiles from \S\ref{sub:2cp} to \S\ref{sub:3cp}
can be understood from Eq.~(\ref{eq:nablaX2}): This equation states
that within the convection zone $\sum_{i=1}^{n-1} \chi_{X_i}
\nabla_{X_i}$ follow the adiabatic gradient, but places no such
restriction on the individual $X$ profiles. Therefore, at the top of
the convection zone, the Fe, O, and Se composition profiles rise
sharply with increasing density because of the correspondingly large
drop in the H-He profile. The burning layer acts as a barrier to
convection, such that the convection zone ends abruptly at the base of
this layer.

Finally, in Fig.~\ref{fig:mcp} we present $X_{\rm H} + X_{\rm He}$
profiles when convection is ignored [such that Eqs.~(\ref{eq:dXdPa1})
and (\ref{eq:dXdPa2}) apply across the entire ocean]. The shape of
these profiles results from the balance between accretion driving the
H-He mixture deeper and nuclear reactions converting H-He into heavier
elements as it travels inward; the profiles can be estimated using
Eqs.~(\ref{eq:taccr}) and (\ref{eq:R3a}): $1 - t_{\rm accr}
R_{3\alpha} \simeq X_1 \Rightarrow X_1 \propto \rho^{5/3}$ for $X_1
\alt 0.1$. In addition to the sharp drop at the top of the convection
zone, the profiles when convection is included differ in that they
fall more slowly with increasing density, due to their larger inward
velocity; since $v_{\rm conv} \propto \rho^{-2/3}$ in the deep ocean,
$X_1 \propto \rho^{3/2}$ for $X_1 \alt 10^{-3}$. The $X_{\rm H} +
X_{\rm He}$ profiles therefore cross at some depth in the ocean; i.e.,
convection causes some hydrogen-helium to mix into deeper
layers. While this mixing will change the amount of burning at each
depth in the convection zone, we do not expect the thermal profile in
this region to change, because the amount of helium in the convection
region is very small (less than 10\% by mass of the total helium in
the star, and less than 1\% of the total mass at any depth in the
convection zone; see Fig.~\ref{fig:mcp}).

\section{Effect of mixing on the thermal profile}
\label{sec:therm}

So far we have taken the ocean to be isothermal, which is a good first
approximation for the bulk of the ocean where thermal conductivity
efficiently transports heat (Bildsten \& Cutler 1995). In this
section, we discuss the effect of convective mixing on the thermal
profile and evaluate the heating associated with the transport of
light elements upwards through the ocean and the steady-state thermal
profile.

When mixing is unimportant, the main contribution to the heat flux in
the ocean comes from energy released by nuclear reactions in the
crust. These reactions release an energy of $\approx 1.7~{\rm MeV}$
per nucleon (e.g., Haensel \& Zdunik 2008), of which $Q_b \sim
0.1$--$1~{\rm MeV}$ per nucleon flows outward through the ocean,
depending on the accretion rate (Brown 2000). This component of the
heat flux can be written as
\be
F_{\rm crust} = Q_b\frac{\dot{m}}{m_p} \,.
\label{eq:Fcrust}
\ee
The temperature gradient required to carry the heat flux is given by
\be
\nabla = \frac{FH_P}{KT} \,,
\ee
where
\be
K = \frac{4ac}{3\rho \kappa_{\rm rad}} + K_{\rm cd}
\ee
is the total thermal conductivity including both radiation and
conduction contributions, $a$ is the radiation constant, $c$ is the
speed of light, $\kappa_{\rm rad}$ is the radiative opacity (e.g.,
Schatz et al. 1999), $K_{\rm cd} = n_ek_B^2T/(m_\star \nu_c)$ is the
thermal conductivity, $m_\star=m_e+E_F/c^2$ is the effective mass of
the electrons, and $\nu_c$ is the collision frequency. In the ocean,
the relevant collision frequency is that between electrons and ions:
$\nu_c = 4e^4 m_\star \Lambda/(3\pi\hbar^3) \langle Z^2
\rangle/\langle Z \rangle$, where $\Lambda$ is the Coulomb logarithm
(Yakovlev \& Urpin 1980; Schatz et al. 1999). For densities $\rho \agt
10^7~{\rm g/cm^3}$ the electrons are degenerate and relativistic, and
the contribution of radiation to $K$ is negligible; in this case we
find
\bal
\nabla \simeq 0.03~\dot{m}_4 {}& \left(\frac{Q_b}{0.1~{\rm MeV}}\right)\left(\frac{T_8}{3}\right)^{-2} \nonumber\\
 {}& \qquad \times \left(\frac{\langle Z^2 \rangle/\langle Z \rangle}{30}\right)\left(\frac{g_{14}}{2.45}\right)^{-1} \,,
\label{eq:delT}
\eal
where in the above we have set $\Lambda=1$. We see therefore that over
most of the ocean a shallow temperature gradient is sufficient to
conduct the heat flux from the crust, and in particular $\nabla \ll
\nabla_{\rm ad} \sim 0.35$ as we assumed in \S\ref{sec:convect}. Note
that this is not true for $\rho \alt 10^7~{\rm g/cm^3}$, where the
electrons are nonrelativistic. In that regime $H_P/K_{\rm cd} \propto
\rho^{-1/3}$, such that at low densities $FH_P/(K_{\rm cd}T) >
\nabla_{\rm ad}$. The radiation contribution to the total $K$ becomes
important at these densities, partially offsetting the effect of a low
$K_{\rm cd}$. Nevertheless, under certain conditions (e.g., $\dot{m}
\ge \dot{m}_{\rm Edd}$ and $T < 10^9~{\rm K}$) $\nabla > \nabla_{\rm
ad}$ at the top of the ocean. It is unclear what happens in that case;
a complete understanding may require a time-dependent calculation (see
\S\ref{sec:discuss}).

Thermal conduction also readily conducts the latent heat away from the
ocean floor. We can estimate the latent heat release using the
equations from Paper I for the free energies of the liquid and solid
states [e.g., equations~(23) and (24) of that paper]; the latent heat
released on freezing is of order $k_BT/\langle A \rangle \approx
0.1~T_8[60/\langle A \rangle]~{\rm keV}$ per nucleon. Much smaller
than $Q_b$, the latent heat is removed with only a small temperature
gradient. This contrasts with, e.g., freezing of solid material in the
Earth's core, in which the temperature gradient required to conduct
the latent heat away is $\nabla > \nabla_{\rm ad}$, so that the latent
heat release drives thermal convection (Stevenson 1981).

When chemical separation drives convection in the ocean, we must also
consider the convective heat flux in addition to the conductive heat
flux. In mixing length theory the heat flux is
\be
\mathbf{F}_{\rm conv} = \frac{\xi}{2} \rho \mathbf{v}_{\rm conv} c_P T(\nabla - \nabla_{\rm ad}) \,,
\ee
where $c_P$ is the heat capacity. Using Eq.~(\ref{eq:vconv3}) for the
convective velocity together with the fact that $\rho v_{\rm
accr}=\dot{m}$, we find
\be
\mathbf{F}_{\rm conv} = -F_{\rm conv}\hat{r}
\ee
with
\be
F_{\rm conv} = \frac{c_PT\dot{m}}{\chi_T} \sum_{i=1}^{n-1} \chi_{X_i}\frac{X_i-X_{i,0}}{X_i} \,.
\ee
Here $\hat{r}$ is the radial unit vector. Note that $F_{\rm conv}>0$
so that the convection transports heat inwards. This is because,
unlike thermally-driven convection, $\nabla < \nabla_{\rm ad}$ so that
a fluid element displaced adiabatically outwards is cooler than its
surroundings at its new location.

\begin{figure}
\begin{center}
\includegraphics[width=\columnwidth]{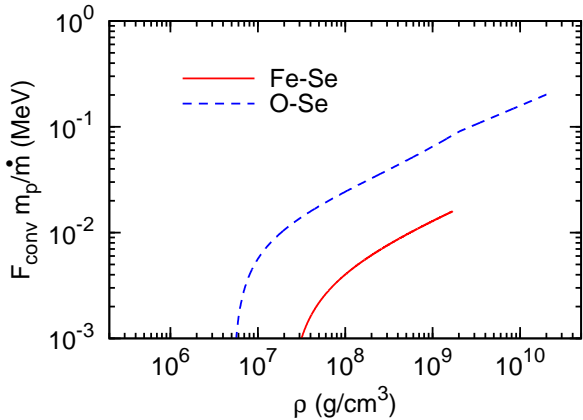}
\end{center}
\caption[The convective flux]
{(Color online) The convective flux for a $\dot{m}=3\times10^4~{\rm
g~cm^{-2}~s^{-1}}$, $T = 3\times10^8$~K ocean composed of ${}^{56}$Fe,
${}^{79}$Se (solid line) and ${}^{16}$O, ${}^{79}$Se (dashed line).}
\label{fig:flux}
\end{figure}

We show the convective flux as $F_{\rm conv} m_p/\dot{m}$ with
$\dot{m} = 3\times10^4~{\rm g~cm^{-2}~s^{-1}}$ in
Fig.~\ref{fig:flux}. In the ocean, the heat capacity is set by the
ions; using the internal energy expansion from DeWitt \& Slattery
(2003) we find $c_P \sim (1$--$4) k_B/[\langle A \rangle m_p]$ giving
$c_PT/m_p \sim (0.1$--$0.6)T_8[60/\langle A \rangle]~{\rm keV}$ per
nucleon, where the range of values is across the depth of the
ocean. Although $c_PT/m_p$ is much smaller than $Q_b$, the convective
flux has an additional factor of $\chi_X/\chi_T \propto
E_F/(k_BT)$. This gives an extra factor of $10$--$100$, so that the
final convective flux is $\sim 10^{-2}$--$10^{-1}~{\rm MeV}$ per
nucleon, which can be comparable to $Q_b$. For the O-Se ocean in
Fig.~\ref{fig:flux}, the convective flux is $\approx 0.2~{\rm MeV}$
per nucleon at the base of the convection zone. For Fe-Se, the smaller
contrast between the heavy and light elements gives a much smaller
flux. For two species, the composition flux is $F_X=\dot m (X-X_0)$,
so that we can write $F_{\rm conv} = F_X c_P T\chi_X/(X\chi_T) = F_X
c_P (\partial \ln T/\partial \ln X|_{P,\rho})$. The convective heat
flux corresponds to the rate of change of internal energy given the
flux of composition at each depth [compare equation~(\ref{eq:taccr})
of Montgomery et al. 1999].

Figure~\ref{fig:flux} shows that the inwards convective flux increases
with depth in the ocean $dF_{\rm conv}/dy>0$. This implies that there
is a local cooling at each depth in the ocean, and so while it is a
good approximation to assume the ocean is isothermal when calculating
the convective velocity and composition profile of the mixed zone, in
fact the isothermal models we presented in \S\ref{sec:convect} are not
in a thermal steady-state. To see what the thermal steady-state must
look like, we write down the entropy equation
\be
T\frac{Ds}{Dt} = \frac{dF}{dy} + \epsilon \,,
\label{eq:entropy}
\ee
where $s$ is the specific entropy, $\epsilon$ is the sum of all
sources and sinks of heat (latent heat release, nuclear reactions,
neutrino cooling, etc.), and $D/Dt = d/dt + \dot{m} d/dy$ is the total
derivative for a fluid element. The heat flux $F$ is the sum of
convective and conductive heat fluxes. In steady state ($ds/dt = 0$)
the left hand side of Eq.~(\ref{eq:entropy}) is given by (e.g., Brown
\& Bildsten 1998)
\be
T\frac{Ds}{Dt} = \frac{c_PT\dot{m}}{y} \left(\nabla - \nabla_{\rm ad} + \frac{1}{\chi_T} \sum_{i=1}^{n-1} \chi_{X_i} \nabla_{X_i} \right) \,,
\ee
much smaller than the $dF/dy$ terms on the right hand side since in
the convective region $\sum_{i=1}^{n-1} \chi_{X_i} \nabla_{X_i}
\approx \chi_T (\nabla_{\rm ad}-\nabla)$ [Eq.~(\ref{eq:nablaX2})]
(the convection zone is very close to adiabatic). Neglecting any
contributions to $\epsilon$ in the ocean, e.g. nuclear reactions or
the small latent heat release at the boundary, we must have $dF/dy=0$
in steady state, or an outwards conductive flux $F_{\rm cond} = F_{\rm
crust}+F_{\rm conv}$. The picture we arrive at is that starting with
an isothermal ocean, the inwards convection of heat steepens the
temperature gradient until the outwards conductive flux balances the
convective flux.

\begin{figure}
\begin{center}
\includegraphics[width=\columnwidth]{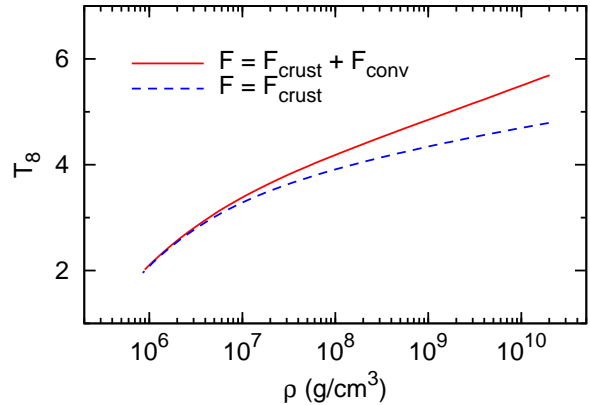}
\end{center}
\caption[The thermal profile]
{(Color online) The thermal profile for a $\dot{m}=3\times10^4~{\rm
g~cm^{-2}~s^{-1}}$ ocean composed of ${}^{16}$O, ${}^{79}$Se and
with $T = 2\times10^8~{\rm K}$ at $\rho = 8\times10^5~{\rm g/cm^3}$,
both when the convective flux is included in the total heat flux $F$
(solid line), and when it is ignored (i.e., the total heat flux is due
only to the outward flux from the crust; dashed line).}
\label{fig:temp}
\end{figure}

We see therefore that the effect of the mixing is to add a
contribution to the conductive flux in the deep ocean of as much as
$0.2$~MeV per nucleon for an O-Se mixture. At high accretion rates
$\dot{m} \agt 0.1~\dot{m}_{\rm Edd}$, this is comparable to or larger
than the flux from the crust $Q_b \sim 0.1$~MeV per nucleon (Brown
2000; at lower accretion rates $\dot{m} \sim 0.01~\dot{m}_{\rm Edd}$,
a larger fraction of the heat released in the crust flows outwards,
and $Q_b \sim 1$~MeV). We show in Fig.~\ref{fig:temp} the thermal
profile with and without the convective heating included, for an O-Se
mixture with $\dot{m}_4=3$ and $T = 2\times10^8~{\rm K}$ at $\rho =
8\times10^5~{\rm g/cm^3}$.


\section{Discussion}
\label{sec:discuss}

We have explored the consequences of chemical separation in the ocean
of accreting neutron stars. Given the rapid timescales for nucleation,
growth, and sedimentation of solid particles, fluid elements that are
lighter than their surrounding are continually being released at the
base of the neutron star ocean. Using a mixing length model for
convection, we modeled the resulting mixing zone. The conclusions are
that the entire ocean is mixed, with a composition gradient $\nabla_X
\approx \chi_T \nabla_{\rm ad}/\chi_X \sim 100k_BT/[\langle Z \rangle
E_F]$ that is very shallow in the bulk of the ocean where the
electrons are relativistically degenerate ($E_F \gg k_BT$). The
composition profile is therefore rather uniform and enhanced in light
elements compared to the composition produced by nuclear burning of
the accreted light elements at the top of the ocean. For example, for
the largest charge ratio we considered, a mixture of O and Se, only
2\% oxygen by mass was added to the ocean by nuclear burning, but the
ocean itself is enriched in steady state to almost 40\% by mass of
oxygen (Fig.~\ref{fig:tcp}), set by the phase diagram for the O-Se
mixture.

Accompanying the outwards transport of light elements is an inwards
transport of heat. In \S\ref{sec:therm}, we argued that the ocean
will evolve to a steady-state in which the convective heat flux is
balanced by an outwards conductive flux. The effect of chemical
separation is therefore to add a contribution to the outwards
conductive heat flux in the ocean, steepening the temperature
gradient. For the O-Se case, we find a heat flux of 0.2~MeV per
nucleon at the base of the ocean, comparable to the flux expected from
the crust. The heating is smaller for an Fe-Se ocean, which has a
smaller contrast in composition between the two species. However, only
a small amount of oxygen in the material entering the ocean is
required to generate significant heating. In the O-Se case we
considered, only 2\% of the incoming mass is oxygen, compared to 40\%
following enrichment. Therefore, in an ocean consisting of many
species, even a small amount of light element entering the ocean could
lead to significant heating through its enrichment. We are currently
investigating the phase diagrams of multicomponent mixtures to assess
the level of heating in that case.

An immediate application of our results is to superbursts. Following
Horowitz et al. (2007) and Paper I, we chose oxygen as the light
element for our example in this paper, but we have calculated models
with carbon as the light element with similar results. Carbon
enrichment through chemical separation at the base of the ocean, and
subsequent upward mixing, could bring the carbon mass fraction $X_C$
at the superburst ignition depth to within the required range $X_C
\approx 0.1$--$0.2$ inferred from observations. Similarly, the heating
associated with chemical separation will help to bring the temperature
of the ocean up to the required ignition temperatures of $5$--$6\times
10^8~{\rm K}$. Questions still remain, however, such as whether this
picture could match the observed recurrence times. The hotter ocean
could also drive a heat flux into the outer crust, explaining the
inverted temperature gradient found by Brown \& Cumming (2009) in
cooling transients. Time-dependent calculations of the evolution of
the ocean are in progress to address these issues.

Compositionally-driven convection is important in cooling white dwarfs
as well. The main difference between convection in the cores of
cooling white dwarfs and in the oceans of accreting neutron stars is
that in the white dwarf case a given mass element crosses the liquidus
curve and crystallizes because the star cools, not because the element
is pushed inwards. The similarities between these two systems are
numerous. For example, the convectively stable composition gradient in
the white dwarf interior is $\nabla_{X,\rm WD} \approx 0.03$, nearly
the same as the gradient in the neutron star ocean; the white dwarf
crystallization rate, $\dot{m}_{\rm WD} \sim 0.1~{\rm
g~cm^{-2}~s^{-1}}$, is also similar (Mochkovitch 1983). Therefore,
we expect that much of our analysis and calculation from the present
paper is applicable to the white dwarf case.

Another situation in which compositionally-driven convection is
important is in the Earth's core, in which chemical separation at the
boundary between the liquid outer core and solid inner core drives
convection in the outer core and provides an energy source for the
Earth's dynamo (Stevenson 1981). There are interesting differences
between the Earth's core and the neutron star ocean, however. First,
the thermal gradient $\nabla$ required to carry away the latent heat
from the Earth's inner-outer core boundary is significant, and exceeds
$\nabla_{\rm ad}$, so that the latent heat drives thermal
convection. Second, in the Earth's core $\nabla_{\rm ad} < \nabla_L$,
where $\nabla_L$ is the liquidus temperature gradient (how the melting
temperature varies with pressure). This allows the boundary between
the inner and outer core to be at the melting temperature, but the
outer core with an adiabatic profile remains above the melting
temperature. Therefore a fluid thermally convecting outer core is
possible (Loper \& Roberts 1977; Stevenson 1980; the opposite was
proposed by Higgins \& Kennedy 1971 in what became known as the ``core
paradox''). In the neutron star ocean, however, $\nabla_{\rm ad} >
\nabla_L$: Equation~(\ref{eq:gamma}) shows that for fixed $\Gamma_i$,
$P \propto T^4$ so that $\nabla_L = 0.25$; whereas using the results
of DeWitt \& Slattery (2003) to calculate Coulomb corrections we find
$\nabla_{\rm ad} \approx 0.36$ with a weak $\Gamma_i$ dependence near
the melting point. This means that the entire ocean could never be
thermally convective and remain liquid.

The steady-state convective velocity is larger than the accretion
velocity by a factor $(X_b - X_0)/(X_b\nabla_X)$, where $X_b - X_0$ is
the required change in composition across the ocean. Because the
marginally-stable composition gradient in the ocean is very small,
this factor is much greater than unity; in the deep ocean, this factor
is of order 100. However, the corresponding velocity is still
extremely slow such that additional processes that we have not
considered here could change the picture we have put forward. For
example, an extremely weak magnetic field (only about $100~{\rm G}$ at
the base of the ocean) would have an energy density comparable to the
kinetic energy of the flow. The dipole magnetic fields of accreting
neutron stars in low mass X-ray binaries are believed to be $\sim
10^9~{\rm G}$ and stronger horizontal field components would not be
surprising, e.g. due to screening currents (Cumming, Zweibel, \&
Bildsten 2001) or winding due to differential rotation and subsequent
instabilities (Piro \& Bildsten 2007). On a rotating star,
Eddington-Sweet circulation in the ocean would have a timescale $\sim
t_{\rm therm} (\Omega_B/\Omega)^2$ where $\Omega_B$ is the break-up
spin frequency. This timescale could be hundreds of days for a thermal
time of days at the base of the ocean and $\Omega/\Omega_B \approx
0.1$, comparable to the convective turnover time. Further work is
needed to investigate how rotational circulation or magnetic stresses
would affect the models presented here.


The steady state we have been discussing is one in which the unstable
composition gradient in the ocean is balanced by the stable
temperature gradient, such that the ocean is only marginally unstable
to convection. We find, however, that under certain conditions (e.g.,
$\dot{m} \ge \dot{m}_{\rm Edd}$) the top of the ocean is thermally
unstable to convection: the temperature gradient must satisfy $\nabla
> \nabla_{\rm ad}$ in order to conduct away the large flux from the
crust. In this case, either the convective velocity must be very large
at the top of the ocean, or the unstable temperature gradient must be
balanced by a {\it stable} composition gradient. The latter scenario
is inconsistent with the two-component models of \S\ref{sub:2cp}, where
the fraction of the light element increases with depth; but could be
consistent with the multicomponent models of \S\ref{sub:3cp} if the
fraction of hydrogen-helium drops faster than oxygen or iron rises. We
are currently studying the time-dependent case to understand what
happens when $\nabla > \nabla_{\rm ad}$.

In addition to oceans with thermally-driven convection, there are
several other scenarios that are likely to lead to interesting time
dependence of the mixing zone. For example, in transients in
quiescence (see, e.g., Shternin et al. 2007; Brown \& Cumming 2009)
the mixing zone relaxes on a timescale comparable to the convective
turnover time [Eq.~(\ref{eq:tconv})], about a month in the deep
ocean. Mixing also occurs as the star cools after an accretion
episode; the bulk of the ocean will solidify after a few days of
cooling and will undergo chemical separation and further
convection. Either of these effects could lead to late-time energy
release, and provide a further observational test of the conditions in
the ocean.

\acknowledgements
We thank Chuck Horowitz, Sanjay Reddy, and Chris Malone for useful
discussions. We are grateful for support from NSERC and the Canadian
Institute for Advanced Research (CIFAR). AC thanks the Kavli
Insititute for Astronomy and Astrophysics (KIAA) Beijing for
hospitality during completion of this work.

\references

\noindent
Bildsten, L. \& Cumming, A. 1998, \apj, 506, 842

\noindent
Bildsten, L. \& Cutler, C. 1995, \apj, 449, 800

\noindent
Bildsten, L. \& Hall, D. M. 2001, \apj, 549, L219

\noindent
Brown, E. F. 2000, \apj, 531, 988

\noindent
Brown, E. F. \& Bildsten, L. 1998, \apj, 496, 915

\noindent
Brown, E. F., Bildsten, L., \& Chang, P. 2002, \apj, 574, 920

\noindent
Brown, E. F. \& Cumming, A. 2009, \apj, 698, 1020

\noindent
Cooper, R. L. \& Bildsten, L. 2008, \pre, 77, 056405

\noindent
Cox, J. P. 1980, Theory of Stellar Pulsation (Princeton: Princeton Univ. Press)

\noindent
Cumming, A. \& Bildsten, L. 2001, \apj, 559, L127

\noindent
Cumming, A., Macbeth, J., in~'t Zand, J. J. M., \& Page, D. 2006, \apj, 646, 429

\noindent
Cumming, A., Zweibel, E., \& Bildsten, L. 2001, \apj, 557, 958

\noindent
Daligault, J. 2006a, \prl, 96, 065003

\noindent
Daligault, J. 2006b, \pre, 73, 056407

\noindent
Daligault, J. \& Murillo, M. S. 2005, \pre, 71, 036408

\noindent
DeWitt, H. \& Slattery, W. 2003, Plasma Phys., 43, 279

\noindent
Donk\'o, Z., \& Ny\'iri, B. 2000, Phys.~Plasmas, 7, 45

\noindent
Gupta, S., Brown, E. F., Schatz, H., M\"oller, P., \& Kratz, K.-L. 2007, \apj, 662, 118

\noindent
Hansen, C. J. \& Kawaler, S. D. 1994, Stellar Interiors: Physical Principles, Structure and Evolution (New York: Springer Verlag)

\noindent
Hansen, J.-P., McDonald, I. R., \& Pollock, E. L. 1975, \pra, 11, 1025

\noindent
Higgins, G. \& Kennedy, G. C. 1971, J. Geophys. Res., 76, 1870

\noindent
Horowitz, C. J., Berry, D. K., \& Brown, E. F. 2007, \pre, 75, 066101

\noindent
Horowitz, C. J., Hughto, J., Schneider, A. S., \& Berry, D. K. 2010, preprint (arXiv:1009.4248v1)

\noindent
Ichimaru, S., Iyetomi, H., Mitake, S., \& Itoh, N. 1983, \apj, 265, L83

\noindent
in~'t Zand, J. J. M., Kuulkers, E., Verbunt, F., Heise, J., \& Cornelisse, R. 2003, \aap, 411, L487

\noindent
in~'t Zand, J. J. M., Cumming, A., van der Sluys, M. V., Verbunt, F., \& Pols, O. R. 2005, \aap, 441, 675

\noindent
Keek, L., in~'t Zand, J. J. M., Kuulkers, E., Cumming, A., Brown, E. F., \& Suzuki, M. 2008, \aap, 479, 177

\noindent
Kelton, K. F., Greer, A. L., \& Thompson, C. V. 1983, \jcp, 79, 6261

\noindent
Kippenhahn, R. \& Weigert, A. 1994, Stellar Structure and Evolution (Springer-Verlag Berlin Heidelberg New York)

\noindent
Kuulkers, E., int 't Zand, J. J. M., Homan, J., van Straaten, S., Altamirano, D., \& van der Klis, M. 2004, in AIP Conf. Proc. 714, X-Ray Timing 2003: Rossi and Beyond, ed. P Kaaret, F. K. Lamb, \& J. H. Swank (Melville: AIP), 257

\noindent
Loper, D. E. \& Roberts, P. H. 1977, Geophys. Astrophys. Fluid Dynamics, 9, 289

\noindent
Medin, Z. \& Cumming, A. 2010, \pre, 81, 036107

\noindent
Mochkovitch, R. 1983, \aap, 122, 212

\noindent
Montgomery, M. H., Klumpe, E. W., Winget, D. E., \& Wood, M. A. 1999, \apj, 525, 482

\noindent
Paczy\'nski, B. 1983, \apj, 267, 31

\noindent
Peng, F., Brown, E. F., \& Truran, J. W. 2007, \apj, 654, 1022

\noindent
Piro, A. L., \& Bildsten, L. 2005, \apj, 619, 1054

\noindent
Piro, A. L. \& Bildsten, L. 2007, \apj, 663, 1252

\noindent
Potekhin, A. Y. \& Chabrier, G. 2000, \pre, 62, 8554

\noindent
Schatz, H., Aprahamian, A., Barnard, V., Bildsten, L., Cumming, A., Ouellette, M., Rauscher, T., Thielemann, F.-K., \& Wiescher, M. 2001, \prl, 86, 3471

\noindent
Schatz, H., Bildsten, L., Cumming, A., \& Ouellette, M. 2003, Nucl. Phys. A, 718, 247

\noindent
Shternin, P. S., Yakovlev, D. G., Haensel, P., \& Potekhin, A. Y. 2007, \mnras, 382, 43

\noindent
Stevenson, D. J. 1980, Physics of the Earth and Planetary Interiors, 22, 42

\noindent
Stevenson, D. J. 1981, Science, 214, 611

\noindent
Strohmayer, T. E., \& Brown, E. F. 2002, \apj, 566, 1045

\noindent
Vehkam\"aki, H. 2006, Classical nucleation theory in multicomponent systems (Springer-Verlag: Berlin Heidelberg)

\noindent
Woosley, S. E., Heger, A., Cumming, A., Hoffman, R. D., Pruet, J., Rauscher, T., Fisker, J. L., Schatz, H., Brown, B. A., \& Wiescher, M. 2004, \apjs, 151, 75

\noindent
Yakovlev, D. G. \& Urpin, V. A. 1980, \sovast, 24, 303

\end{document}